\documentclass[aps,pra,twocolumn,nobibnotes,superscriptaddress,nofootinbib,10pt]{revtex4-1} 
\usepackage[latin1]{inputenc}
\usepackage{amsmath,amssymb}
\usepackage{mathrsfs} 
\usepackage{graphicx,color,colortbl}
\usepackage{rotating,array,tabularx,booktabs}
\usepackage{varwidth,xcolor}
\usepackage{placeins}
\usepackage{siunitx}
\usepackage{layouts}

\DeclareGraphicsExtensions{.pdf,.PDF,.png,.PNG}
\allowdisplaybreaks

\newcommand{\norm}[1]{\lVert#1\rVert}%
\newcommand{\uu}{\hat{n}}%
\newcommand{\ww}{\vec{u}}%
\newcommand{\ws}{u}%

\begin{document}

\title{Dependence of the acoustic propulsion of nano- and microcones on their orientation and aspect ratio}

\author{Johannes Vo\ss{}}
\affiliation{Institut f\"ur Theoretische Physik, Center for Soft Nanoscience, Westf\"alische Wilhelms-Universit\"at M\"unster, 48149 M\"unster, Germany\\$^{*}$\,Corresponding author: raphael.wittkowski@uni-muenster.de}

\author{Raphael Wittkowski}
\affiliation{Institut f\"ur Theoretische Physik, Center for Soft Nanoscience, Westf\"alische Wilhelms-Universit\"at M\"unster, 48149 M\"unster, Germany\\$^{*}$\,Corresponding author: raphael.wittkowski@uni-muenster.de}

\begin{abstract}
Recent research revealed the orientation-dependent propulsion of a cone-shaped colloidal particle that is exposed to a planar traveling ultrasound wave. Here, we extend the previous research by considering nano- and microcones with different aspect ratios and studying how the propulsion of a particle depends on its orientation and aspect ratio. We also study how the orientation-averaged propulsion of a cone-shaped particle, which corresponds to an isotropic ultrasound field, depends on its aspect ratio and identify an aspect ratio of $1/2$ where the orientation-averaged propulsion is particularly strong. To make our simulation results easier reusable for follow-up research, we provide a corresponding simple analytic representation. 
\end{abstract}
\maketitle

\section{Introduction}
After the discovery of ultrasound-propelled nano- and microparticles in 2012 \cite{WangCHM2012}, a decade of intensive and still rapidly growing research on this type of artificial motile particles followed \cite{WangCHM2012,GarciaGradillaEtAl2013,AhmedEtAl2013,NadalL2014,WuEtAl2014,WangLMAHM2014,GarciaGradillaSSKYWGW2014,BalkEtAl2014,AhmedGFM2014,WangDZSSM2015,EstebanFernandezdeAvilaMSLRCVMGZW2015,WuEtAl2015a,WuEtAl2015b,RaoLMZCW2015,ZhouZWW2017,KimGLZF2016,EstebanEtAl2016,SotoWGGGLKACW2016,AhmedWBGHM2016,KaynakONNLCH2016,UygunEtAl2017,EstebanFernandezEtAl2017,RenZMXHM2017,CollisCS2017,ZhouYWDW2017,ChenEtAl2018,HansenEtAl2018,SabrinaTABdlCMB2018,AhmedBJPDN2016,Zhou2018,WangGWSGXH2018,EstebanEtAl2018,LuSZWPL2019,TangEtAl2019,QualliotineEtAl2019,GaoLWWXH2019,KaynakONLCH2017,RenEtAl2019,AghakhaniYWS2020,LiuR2020,VossW2020,ValdezLOESSWG2020,DumyJMBGMHA2020,VossW2021,VossW2022acousticb,VossW2022orientation,VossW2022acoustica,VossW2022propulsion,VossW2022microspinner,MohantyEtAl2021}. 
So far, the research was mostly experimental \cite{WangCHM2012,GarciaGradillaEtAl2013,AhmedEtAl2013,WuEtAl2014,WangLMAHM2014,GarciaGradillaSSKYWGW2014,BalkEtAl2014,AhmedGFM2014,WangDZSSM2015,EstebanFernandezdeAvilaMSLRCVMGZW2015,WuEtAl2015a,WuEtAl2015b,EstebanEtAl2016,SotoWGGGLKACW2016,AhmedWBGHM2016,KaynakONNLCH2016,UygunEtAl2017,KaynakONLCH2017,EstebanFernandezEtAl2017,RenZMXHM2017,ZhouYWDW2017,HansenEtAl2018,SabrinaTABdlCMB2018,AhmedBJPDN2016,ZhouZWW2017,WangGWSGXH2018,EstebanEtAl2018,Zhou2018,TangEtAl2019,QualliotineEtAl2019,GaoLWWXH2019,RenEtAl2019,AghakhaniYWS2020,LiuR2020,ValdezLOESSWG2020,DumyJMBGMHA2020}, but it includes also computer simulations \cite{AhmedBJPDN2016,SabrinaTABdlCMB2018,Zhou2018,TangEtAl2019,VossW2020,VossW2021,VossW2022acousticb,VossW2022orientation,VossW2022acoustica,VossW2022propulsion,VossW2022microspinner} and analytical approaches \cite{NadalL2014,CollisCS2017}. 
Reasons for the intensive investigation of the particles are that the acoustic propulsion is biocompatible \cite{WangLWXLGM2019,OuEtAl2020} and fuel-free \cite{XuGXZW2017} and that it provides a simple way to supply the particles permanently with energy \cite{XuGXZW2017,WangLWXLGM2019}.  
These special properties make the particles relevant for future applications in, e.g., medicine \cite{LiEFdAGZW2017,PengTW2017,SotoC2018,WangGZLH2020,WangZ2021} and materials science \cite{JunH2010,McdermottKDKWSV2012,KuemmelSLVB2015,VanderMeerDF2016,JeanneretPKP2016,NeedlemanD2017,WangDZLGYLWM2019,RamananarivoDP2019,FratzlFKS2021}. 

Nevertheless, much further research is still needed to complete the step from the discovery of these particles to their envisaged practical application. 
For example, nearly all previous studies considered particles in a standing ultrasound wave \cite{WangCHM2012,AhmedEtAl2013,GarciaGradillaEtAl2013,GarciaGradillaSSKYWGW2014,WangLMAHM2014,AhmedGFM2014,BalkEtAl2014,WuEtAl2015a,Kiristi2015,RaoLMZCW2015,SotoWGGGLKACW2016,AhmedWBGHM2016,EstebanEtAl2016,ZhouYWDW2017,ZhouZWW2017,EstebanFernandezEtAl2017,UygunEtAl2017,XuXZ2017,Zhou2018,SabrinaTABdlCMB2018,WangGWSGXH2018,EstebanEtAl2018,QualliotineEtAl2019,TangEtAl2019,LuSZWPL2019,BhuyanDBSGB2019,DumyJMBGMHA2020,LiuR2020,ValdezLOESSWG2020,McneillSWOLNM2021}, since this simplifies the experimental setup, whereas a traveling ultrasound wave would be much more relevant with respect to future applications of these particles \cite{VossW2020,VossW2021,VossW2022acousticb,VossW2022orientation,VossW2022acoustica,VossW2022propulsion,VossW2022microspinner,AhmedBJPDN2016}. 
Furthermore, in a standing wave, the particles' orientations typically align with a nodal plane of the ultrasound field so that most of the existing studies on ultrasound-propelled particles considered only particles with particular orientations relative to the ultrasound wave \cite{WangCHM2012,AhmedEtAl2013,GarciaGradillaEtAl2013,GarciaGradillaSSKYWGW2014,WangLMAHM2014,AhmedGFM2014,BalkEtAl2014,WuEtAl2015a,Kiristi2015,RaoLMZCW2015,SotoWGGGLKACW2016,AhmedWBGHM2016,EstebanEtAl2016,ZhouYWDW2017,ZhouZWW2017,EstebanFernandezEtAl2017,UygunEtAl2017,XuXZ2017,Zhou2018,SabrinaTABdlCMB2018,WangGWSGXH2018,EstebanEtAl2018,QualliotineEtAl2019,TangEtAl2019,LuSZWPL2019,BhuyanDBSGB2019,DumyJMBGMHA2020,LiuR2020,ValdezLOESSWG2020,McneillSWOLNM2021}, whereas particles that can rotate in all directions are much more application-relevant \cite{WuEtAl2015b,VossW2022orientation,NitschkeW2021}. 
 
In a few recent studies \cite{VossW2020,VossW2021,VossW2022acousticb,VossW2022orientation,VossW2022acoustica}, cone-shaped nano- and microparticles in a traveling ultrasound wave have been investigated. It was found that, compared to particles with other shapes, cone-shaped particles exhibit a particularly efficient acoustic propulsion \cite{VossW2020}. 
Reference \cite{VossW2022orientation} considered also different orientations of the particles relative to the propagation direction of the ultrasound wave. It found that the propulsion of the particles has a strong dependence on the particles' orientation.
This study, however, considered only the special case of cone-shaped particles with aspect ratio $\chi=1$, while an understanding of the orientation-dependent propulsion also for other particle shapes will be crucial for their future applications. 

In the present article, we, therefore, continue the previous research by studying how the orientation-dependent propulsion of nano- and microcones by a traveling ultrasound wave changes with the aspect ratio of the particles. 
We also study how the flow field that is generated around the particles depends on their orientation and aspect ratio. Moreover, we consider the orientation-averaged propulsion of the particles, which corresponds to exposing them to an isotropic (i.e., orientation-independent) ultrasound field, such as a superposition of ultrasound waves with different orientations, and study how this propulsion depends on the particles' aspect ratio. 
Our investigation is based on direct acoustofluidic simulations, where the orientation and aspect ratio of the particles can be varied much easier than in experiments. 
To facilitate the reuse of our simulation results by future studies, we provide also a simple analytic representation of these results.

\section{\label{methods}Methods}
We adopt the methodology of Ref.\ \cite{VossW2020}, which is well established for acoustically actuated particles and has been proven to be successful.

\subsection{Setup}
\begin{figure}[htb]
\centering
\includegraphics[width=\linewidth]{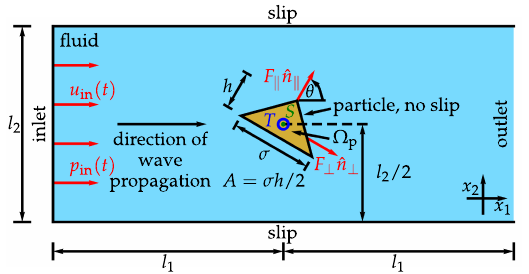}%
\caption{\label{fig:fig1}Setup for the simulations.}
\end{figure}

Figure \ref{fig:fig1} gives an overview of the system that is studied in the present work.  
The simulation domain is rectangular with width $2l_1$ (parallel to the $x_1$ axis) and height $l_2=\SI{200}{\micro\metre}$ (parallel to the $x_2$ axis). It contains a particle with a triangular cross-section, with diameter $\sigma=\sqrt{2A/\chi}$, height $h=\sigma\chi$, cross-section area $A=\sigma h/2=\SI{0.25}{\micro\metre^2}$, and variable height-to-diameter ratio $\chi=h/\sigma\in\{0.5,2,4\}$, that is surrounded by a fluid. 
This fixed particle shape is analogous to Ref.\ \cite{VossW2021}.
We define the particle's position by its center of mass $\mathrm{S}$, and the particle's orientation by the orientational unit vector $\uu_\parallel$ that is parallel to the vector pointing from $\mathrm{S}$ to the tip of the particle.
For completeness, we also introduce an orientational unit vector $\uu_\perp$ that is perpendicular to $\uu_\parallel$.
With a polar angle $\theta$, the orientational unit vectors can be parameterized as 
\begin{align}
\uu_\parallel(\theta) &= (\cos(\theta),\sin(\theta))^\mathrm{T},\\
\uu_\perp(\theta) &= (\sin(\theta),-\cos(\theta))^\mathrm{T}.
\end{align}
The particle's position shall coincide with the geometric center of the rectangular simulation domain, and the particle's orientation $\theta$ is varied from $\theta=0$, where $\uu_\parallel$ is parallel to the $x_1$ axis and $\uu_\perp$ is antiparallel to the $x_2$ axis, to $\theta=\pi$, where $\uu_\parallel$ is antiparallel to the $x_1$ axis and $\uu_\perp$ is parallel to the $x_2$ axis. 

Our assumption that the position of the particle is fixed corresponds to a particle with infinite mass density. This can be seen as an upper limit for the propulsion of a particle made of a material with a high mass density such as gold. 
For lower mass densities $\rho_\mathrm{p}$, lower propulsion speeds can be expected.
Our results can be transferred to particles with a lower mass density by multiplying the obtained propulsion speeds by the factor $(1-\rho_0/\rho_\mathrm{p})$ \cite{NadalL2014}, where $\rho_0$ denotes the mass density of the fluid.

We choose water as an application-relevant fluid. Initially, it shall be at standard temperature $T_0=\SI{293.15}{\kelvin}$, at standard pressure $p_0=\SI{101325}{\pascal}$, and at rest (vanishing velocity field $\ww_0=\vec{0}\,\SI{}{\metre\,\second^{-1}}$). 
Then, its mass density is $\rho_0=\SI{998}{\kilogram\,\metre^{-3}}$, its shear viscosity is $\nu_\mathrm{s}=\SI{1.002}{\milli\pascal\,\second}$, its bulk viscosity is $\nu_\mathrm{b}=\SI{2.87}{\milli\pascal\,\second}$, and its sound velocity is $c_\mathrm{f}=\SI{1484}{\metre\,\second^{-1}}$. 

A planar traveling ultrasound wave with frequency $f=\SI{1}{\MHz}$ and wavelength $\lambda=c_\mathrm{f}/f=\SI{1.484}{\milli \metre}$ shall enter the system at the left edge (inlet) of the fluid domain, propagate in positive $x_1$ direction, interact with the particle, and finally be able to leave the system at its right edge (outlet).
To initiate the entering ultrasound wave at the inlet, we prescribe the time-dependent velocity $\ws_{\mathrm{in}}(t)=\Delta u \sin(2\pi f t)$ and the time-dependent pressure $p_{\mathrm{in}}(t)=\Delta p \sin(2\pi f t)$ there, where $t$ denotes time, $\Delta u=\Delta p / (\rho_0 c_{\mathrm{f}})$ is the velocity amplitude, and $\Delta p=\SI{10}{\kilo\pascal}$ is the pressure amplitude. 
Since the ultrasound wave shall not be damped when it propagates parallel to the lower and upper edges of the fluid domain with length $2l_1=\lambda/2$, we choose slip boundary conditions at these edges. 
In contrast, for an appropriate interaction of the ultrasound with the particle, we choose no-slip boundary conditions at the boundary of the particle domain $\Omega_\mathrm{p}$. 

When the ultrasound wave with acoustic energy density $E=\Delta p^2/(2 \rho_0 c_{\mathrm{f}}^2) =\SI{22.7}{\milli\joule\,\metre^{-3}}$ interacts with the particle, it exerts a propulsion force and a propulsion torque on it. 
We are interested in the time-averaged stationary propulsion forces $F_\parallel$ and $F_\perp$ that act on $\mathrm{S}$ in the direction parallel to $\uu_\parallel$ and $\uu_\perp$, respectively, and the time-averaged stationary propulsion torque $T$ that tends to rotate the particle about $\mathrm{S}$ in the planar simulation domain.

\subsection{Parameters}
The parameters of the system that are relevant for our simulations and the values assigned to these parameters are summarized in Tab.\ \ref{tab:Parameters}.
\begin{table*}[htb]
\caption{\label{tab:Parameters}Relevant parameters of the system and the assigned values, which follow Ref.\ \cite{VossW2020}.}%
\centering
\begin{ruledtabular}
\begin{tabular}{llll}%
\textbf{Name} & \textbf{Symbol} & \textbf{Value} & \textbf{Remark}\\
\hline
Particle cross section area  & $A$ & $\SI{0.25}{\micro\metre^2}$ & \\
Particle height-to-diameter ratio & $\chi=h/\sigma$ & $\SI{0.5}{}$,\,$\SI{1}{}$,$\SI{2}{}$,\,$\SI{4}{}$ & \\
Particle diameter & $\sigma$ & $\sqrt{2A/\chi}$ & \\
Particle height & $h$ & $\sigma\chi$ & \\
Particle orientation angle & $\theta$ & $0$-$\pi$ & \\
Sound frequency & $f$ & $\SI{1}{\mega\hertz}$ & \\
Speed of sound & $c_\mathrm{f}$ & $\SI{1484}{\metre\,\second^{-1}}$ & Corresponds to $T_0,p_0$ \\
Time period of sound & $\tau=1/f$ & $\SI{1}{\micro\second}$ & \\
Wavelength of sound & $\lambda=c_\mathrm{f}/f$ & $\SI{1.484}{\milli\metre}$ & \\
Temperature of fluid & $T_0$ & $\SI{293.15}{\kelvin}$ & \\
Mean mass density of fluid & $\rho_0$ & $\SI{998}{\kilogram\,\metre^{-3}}$ & Corresponds to $T_0,p_0$ \\
Mean pressure of fluid & $p_{0}$ & $\SI{101325}{\pascal}$ & \\
Initial velocity of fluid & $\ww_{0}$ & $\vec{0}\,\SI{}{\metre\,\second^{-1}}$ & \\
Sound pressure amplitude & $\Delta p$ & $\SI{10}{\kilo\pascal}$ & \\
Acoustic energy density & $E=\Delta p^2/(2 \rho_0 c_{\mathrm{f}}^2)$ & $\SI{22.7}{\milli\joule\,m^{-3}}$ & \\
Shear/dynamic viscosity of fluid & $\nu_{\mathrm{s}}$ & $\SI{1.002}{\milli\pascal\,\second}$ & Corresponds to $T_0,p_0$ \\
Bulk/volume viscosity of fluid & $\nu_{\mathrm{b}}$ & $\SI{2.87}{\milli\pascal\,\second}$ & Interpolated from Tab.\ 1 in Ref.\ \cite{HolmesPP2011} for $T_0,p_0$ \\
Inlet-particle/particle-outlet distance & $l_1$ & $\lambda/4$ & \\
Inlet length & $l_2$ & $\SI{200}{\micro\metre}$ & \\
Mesh-cell size & $\Delta x$ & $\SI{15}{\nano \metre}$-$\SI{1}{\micro \metre}$ & \\
Time-step size & $\Delta t$ & $1$-$\SI{10}{\pico \second}$ & \\
Simulation start time & $t_0$ & $0$ & \\
Simulation duration & $t_{\mathrm{max}}$ & $\geqslant 500\tau$ & \\
Euler number & $\mathrm{Eu}$ & $\SI{2.2}{\cdot10^5}$ & \\
Helmholtz number & $\mathrm{He}$ & $\SI{3.37}{\cdot10^{-4}}$ & \\
Bulk Reynolds number &  $\mathrm{Re}_\mathrm{b}$ & $\SI{1.17}{\cdot10^{-2}}$ & \\
Shear Reynolds number &  $\mathrm{Re}_\mathrm{s}$ & $\SI{3.36}{\cdot10^{-2}}$ & \\
Particle Reynolds number &  $\mathrm{Re}_\mathrm{p}$ & $<\SI{3}{\cdot10^{-7}}$ & 
\end{tabular}%
\end{ruledtabular}%
\end{table*}
With these parameter values, the outer boundaries of the fluid domain are so far away from the particle that finite-size effects can be neglected.

\subsection{Acoustofluidic simulations}
We simulate the dynamics of the fluid by numerically solving the standard equations of fluid dynamics (continuity equation, compressible Navier-Stokes equations, linear constitutive equation for pressure) with the finite volume software package OpenFOAM \cite{WellerTJF1998}. 
A nondimensionalization of the equations results in the emergence of four dimensionless characteristic numbers, namely the 
Euler number
\begin{equation}
\mathrm{Eu}=\frac{\Delta p}{\rho_0 \Delta u^2}\approx \SI{2.2}{\cdot10^{5}},
\end{equation}
Helmholtz number
\begin{equation}
\mathrm{He}=\frac{f \sqrt{A}}{c_\mathrm{f}}\approx \SI{3.37}{\cdot 10^{-4}},
\end{equation}
bulk Reynolds number
\begin{equation}
\mathrm{Re}_\mathrm{b}=\frac{\rho_0 \Delta u \sqrt{A}}{\nu_\mathrm{b}}\approx \SI{1.17}{\cdot10^{-3}},
\end{equation}
and shear Reynolds number
\begin{equation}
\mathrm{Re}_\mathrm{s}=\frac{\rho_0 \Delta u \sqrt{A}}{\nu_\mathrm{s}}\approx \SI{3.36}{\cdot10^{-3}}.
\end{equation}
An interpretation of these dimensionless numbers can be found in Ref.\ \cite{VossW2021}. 

To solve the equations for the dynamics of the fluid numerically, we discretize the fluid domain spatially by a structured, mixed rectangle-triangle mesh with about 250,000 cells and a typical cell size $\Delta x$ ranging from $\SI{15}{\nano\metre}$ (near the particle) to $\SI{1}{\micro\metre}$ (far away from the particle).
For discretization in the time domain $[t_0,t_\mathrm{max}]$ with start time $t_0=0$ and end time $t_\mathrm{max} \geqslant 500\tau$, we apply an adaptive time-step method that can vary the time-step size $\Delta t$ in the range from $\SI{1}{\pico\second}$ to $\SI{10}{\pico\second}$, always ensuring that the Courant-Friedrichs-Lewy condition 
\begin{align}
C = c_\mathrm{f} \frac{\Delta t}{\Delta x} < 1 
\end{align}
is met. The typical computational cost for each simulation run is about $36,000$ CPU core hours. 

Since we simulate a particle with a fixed position, the laboratory frame and the particle frame coincide. This is why an Eulerian-grid-based solver such as OpenFOAM can be applied here.
In the chosen coordinate system, the no-slip boundary condition of the particle means that the fluid velocity is prescribed to vanish at the boundary of the particle.

\subsection{Propulsion force and torque}
The acoustofluidic simulations yield the time evolution of the fluid's mass-density field, velocity field, and pressure field. We calculate from these fields the propulsion force and torque that are exerted on the particle. 
In particular, we calculate the time-averaged stationary propulsion forces $F_\parallel$ and $F_\perp$ as well as the time-averaged stationary propulsion torque $T$. 
These quantities are obtained from the stress tensor of the fluid by performing suitable integrals over the particle surface, locally averaging over one time period, and extrapolating towards $t \to \infty$.  
Reference \cite{VossW2020} describes this procedure in detail. 
 
When averaging the propulsion forces or torque over one time period $\tau=\SI{1}{\micro\second}$, the orientation of the particle can be assumed to be constant. This is possible since the particle's rotation within one time period that results from the particle's angular propulsion is negligibly small. As is shown in Section \ref{ResultsPropulsion}, the maximum observed angular propulsion is $\omega_{\mathrm{max}}\approx\SI{2.22}{\second^{-1}}$. Hence, the particle rotates by only $\omega_{\mathrm{max}}\tau\approx 2.22\cdot 10^{-6}$ within one time period.

\subsection{Translational and angular propulsion velocity}
The translational propulsion velocities $v_{\parallel}$ and $v_{\perp}$ parallel and perpendicular to the particle's orientation, respectively, and the angular propulsion velocity $\omega$ can be calculated from $F_\parallel$, $F_\perp$, and $T$ by applying the Stokes law \cite{HappelB1991}.  
A detailed description of this procedure can be found in Ref.\ \cite{VossW2020}. 
We can use the Stokes law since we deal with rigid particles that are propelled by an external field (the ultrasound) exerting propulsion forces and torques on the particles. While for such actuated particles there is a simple mapping from the propulsion forces and torques to the particles' translational and angular propulsion velocities via the Stokes law \cite{LaugaP2009}, the calculation of the propulsion velocities would be much more complicated for nano- or microswimmers, such as microorganisms undergoing shape deformations or chemically propelled Janus particles. Swimmers are not propelled by forces or torques exerted by an external field but by self-generated internal propulsion forces or torques that are associated with a time-dependent particle shape or a flow field with a nonvanishing slip velocity at the particle surface (in contrast to the no-slip boundary condition that applies to our particles). A widely used approach to calculate the propulsion velocities of swimmers is the reciprocal theorem \cite{LaugaP2009}. Examples for other approaches can be found in Refs.\ \cite{LaugaP2009,PicellaM2021}.

Since the procedure requires the hydrodynamic resistance matrix $\boldsymbol{\mathrm{H}}$ of our particle, which can be calculated from the particle's shape, e.g., by using the software \texttt{HydResMat} \cite{VossW2018,VossJW2019}, we here present this matrix for better reproducibility of our results. 
When choosing the particle's center of mass $\mathrm{S}$ as the reference point (see Ref.\ \cite{HappelB1991} for details), specifying the particle orientation as $\theta=\pi/2$, and assuming that the particle has a thickness of $\sigma$ in the third dimension (see Ref.\ \cite{VossW2020} for details), the hydrodynamic resistance matrix of the particle studied in the present work has the form
\begin{align}
\boldsymbol{\mathrm{H}} = \begin{pmatrix}
\mathrm{K_{11}} & 0 & 0 & 0 & 0 & \mathrm{C_{31}}\\
 0 & \mathrm{K_{22}} & 0 & 0 & 0 & 0 \\
 0 & 0 & \mathrm{K_{33}} & \mathrm{C_{13}} & 0 & 0\\
 0 & 0 & \mathrm{C_{13}} & \mathrm{\Omega_{11}} & 0 & 0 \\
 0 & 0 & 0 & 0 & \mathrm{\Omega_{22}} & 0 \\
 \mathrm{C_{31}} & 0 & 0 & 0 & 0 &\mathrm{\Omega_{33}}
\end{pmatrix}.
\end{align}
The values of the nonvanishing elements of $\boldsymbol{\mathrm{H}}$ are given in Tab.\ \ref{tab:resistance} for each aspect ratio of the particle shape that is considered in our study.
\begin{table*}[tb]
\centering
\caption{\label{tab:resistance}Nonzero elements of the hydrodynamic resistance matrix $\boldsymbol{\mathrm{H}}$ of the particle that is studied in the present work (see Fig.\ \ref{fig:fig1}) for different aspect ratios $\chi$ of the particle shape.}
\begin{ruledtabular}
\begin{tabular}{ccccccccc}
$\boldsymbol{\chi}$ & $\mathbf{K_{11}} / \SI{}{\textbf{\micro\metre}}$ & $\mathbf{K_{22}}/ \SI{}{\textbf{\micro\metre}}$ & $\mathbf{K_{33}}/ \SI{}{\textbf{\micro\metre}}$ & $\mathbf{C_{13}}/ \SI{}{\textbf{\micro\metre}^{\mathbf{2}}}$ & $\mathbf{C_{31}}/ \SI{}{\textbf{\micro\metre}^{\mathbf{2}}}$& $\mathbf{\Omega_{11}}/ \SI{}{\textbf{\micro\metre}^{\mathbf{3}}}$ & $\mathbf{\Omega_{22}}/ \SI{}{\textbf{\micro\metre}^{\mathbf{3}}}$ & $\mathbf{\Omega_{33}}/ \SI{}{\textbf{\micro\metre}^{\mathbf{3}}}$ \\
\hline
$\SI{0.5}{}$ & $\SI{8.49}{}$ & $\SI{9.72}{}$ & $\SI{8.14}{}$ & $\SI{-0.14}{}$ & $\SI{0.4}{}$ & $\SI{3}{}$ & $\SI{3.36}{}$ & $\SI{2.63}{}$\\
$\SI{1}{}$ & $\SI{7.74}{}$ & $\SI{7.48}{}$ & $\SI{7.16}{}$ & $\SI{0.05}{}$ & $\SI{-0.11}{}$ & $\SI{1.81}{}$ & $\SI{1.69}{}$ & $\SI{1.73}{}$\\
$\SI{2}{}$ & $\SI{7.67}{}$ & $\SI{6.38}{}$ & $\SI{6.93}{}$ & $\SI{0.35}{}$ & $\SI{-0.66}{}$ & $\SI{1.78}{}$ & $\SI{0.97}{}$ & $\SI{1.97}{}$\\
$\SI{4}{}$ & $\SI{7.89}{}$ & $\SI{5.79}{}$ & $\SI{7.16}{}$ & $\SI{0.8}{}$ & $\SI{-1.31}{}$ & $\SI{2.57}{}$ & $\SI{0.6}{}$ & $\SI{3.11}{}$\\
\end{tabular} 
\end{ruledtabular}%
\end{table*}
This hydrodynamic resistance matrix corresponds to a particle in an unbounded fluid domain. We can use this matrix since in our work the fluid domain is much larger than the particle so that there is no significant influence of the outer boundaries of the simulation domain on the hydrodynamic resistance matrix.

When the propulsion velocities $v_\parallel$ and $v_\perp$ of the particle are known, we can calculate the 
particle Reynolds number
\begin{align}
\mathrm{Re}_\mathrm{p}&=\frac{\rho_0\sqrt{A} }{\nu_\mathrm{s}} \sqrt{v_\parallel^2+v_\perp^2} 
<3\cdot 10^{-7}
\end{align}
that characterizes the motion of the particle through the fluid.
As one can see, this dimensionless number is very small for all orientations and aspect ratios of the particle that are studied in this work. 
This shows that viscous forces dominate inertial forces in particle motion.

\subsection{Error estimation}
The results for $v_\parallel$, $v_\perp$, and $\omega$ are associated with numerical errors. 
We estimate these errors by considering the values of $v_\perp$ and $\omega$ for $\theta=0$ and $\theta=\pi$. 
Since these values should vanish for reasons of symmetry, but will not exactly do so due to numerical inaccuracies of the calculations, we use these deviation values to estimate the numerical errors. 
In particular, we determine the absolute values of $v_\perp$ and $\omega$ and maximize them over both angles. 
We then use the obtained maximum deviation for $v_\perp$ as the estimated error of our results for $v_\parallel$ and $v_\perp$, and we use the maximum deviation for $\omega$ as the estimated error of our results for $\omega$.

\section{\label{results}Results and discussion}
\subsection{Flow field}
\begin{figure*}[tb!]
\centering
\includegraphics[width=\linewidth]{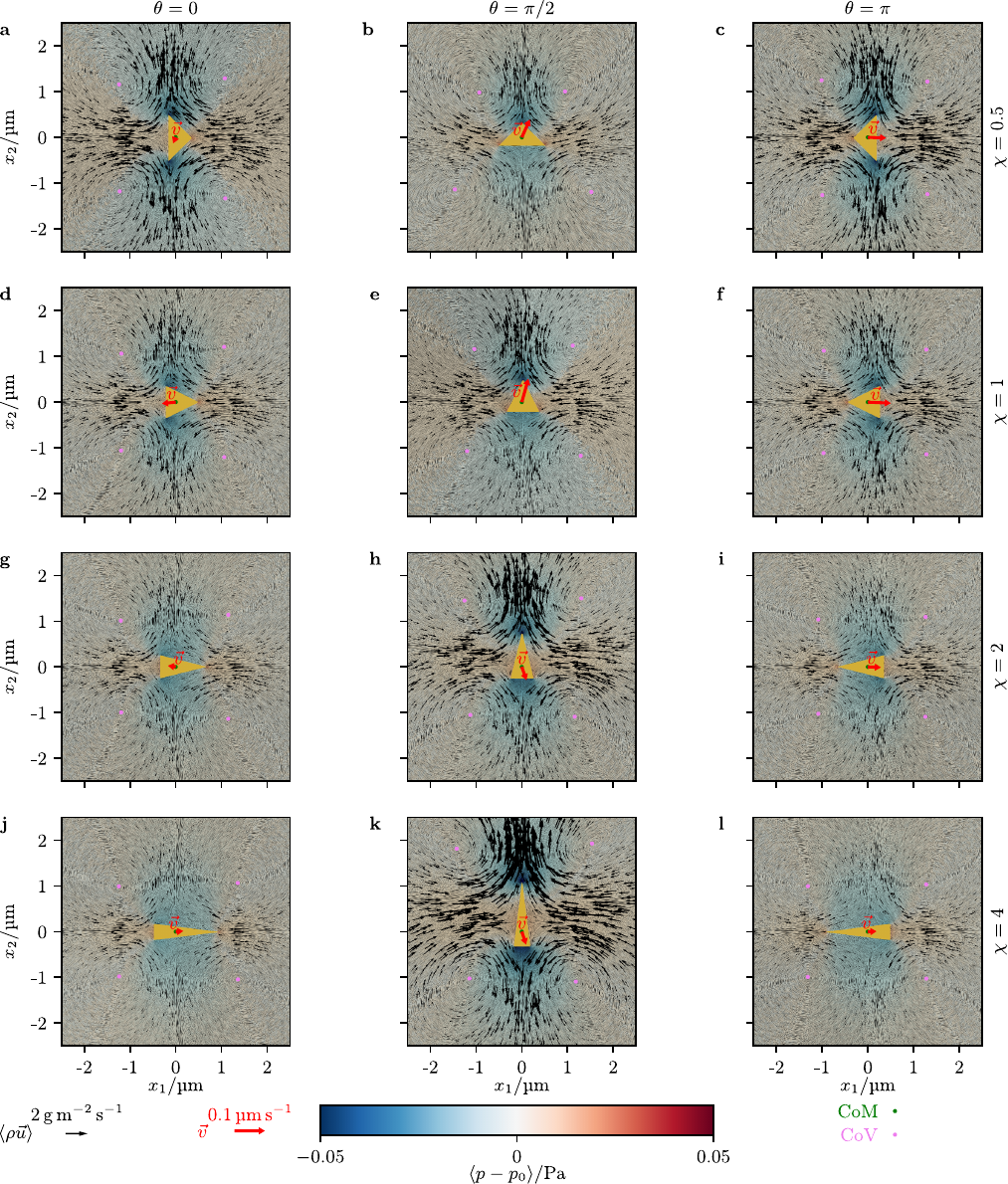}%
\caption{\label{fig:fig2}Time-averaged mass-current density $\langle\rho\ww\rangle$ and reduced pressure $\langle p-p_{0}\rangle$ for different orientations $\theta$ and aspect ratios $\chi$ of the particle.
The center of mass (CoM) of the particle, the centers of vortices (CoV) of the flow field, and the propulsion velocity $\vec{v}$ are also shown.}
\end{figure*}
We first study how the time-averaged stationary flow field around a cone-shaped particle changes for different orientations $\theta$ and aspect ratios $\chi$ of the particle. Figure \ref{fig:fig2} shows our corresponding simulation results. In the background, the time-averaged mass-current density $\langle\rho\ww \rangle$ and reduced pressure $\langle p-p_0\rangle$ of the fluid are shown, where $\rho$ is the mass density, $\ww$ the velocity, and $p$ the pressure of the fluid. The general structure of the flow field is the same for all considered values of $\theta$ and $\chi$. Four vortices, placed at the top left, top right, bottom left, and bottom right relative to the particle, cause the fluid to flow towards the particle from the left and right, and away from the particle above and below it. Therefore, the reduced pressure is negative above and below the particle, whereas it is positive besides the particle. 
When $\chi$ is increased, the strength of the flow field decreases for $\theta=0$ and $\theta=\pi$ (i.e., an orientation parallel or antiparallel to the ultrasound wave), but it increases for $\theta=\pi/2$ (i.e., an orientation perpendicular to the ultrasound wave). 
A possible reason for this behavior is a dependence of the strength of the flow field on the effective cross section of the particle perpendicular to the ultrasound wave. This effective cross section decreases for $\theta=0$ and $\theta=\pi$ but increases for $\theta=\pi/2$ when $\chi$ is increased.

\subsection{\label{ResultsPropulsion}Dependence of acoustic propulsion on orientation and aspect ratio}
\begin{figure*}[tb]
\centering
\includegraphics[width=\linewidth]{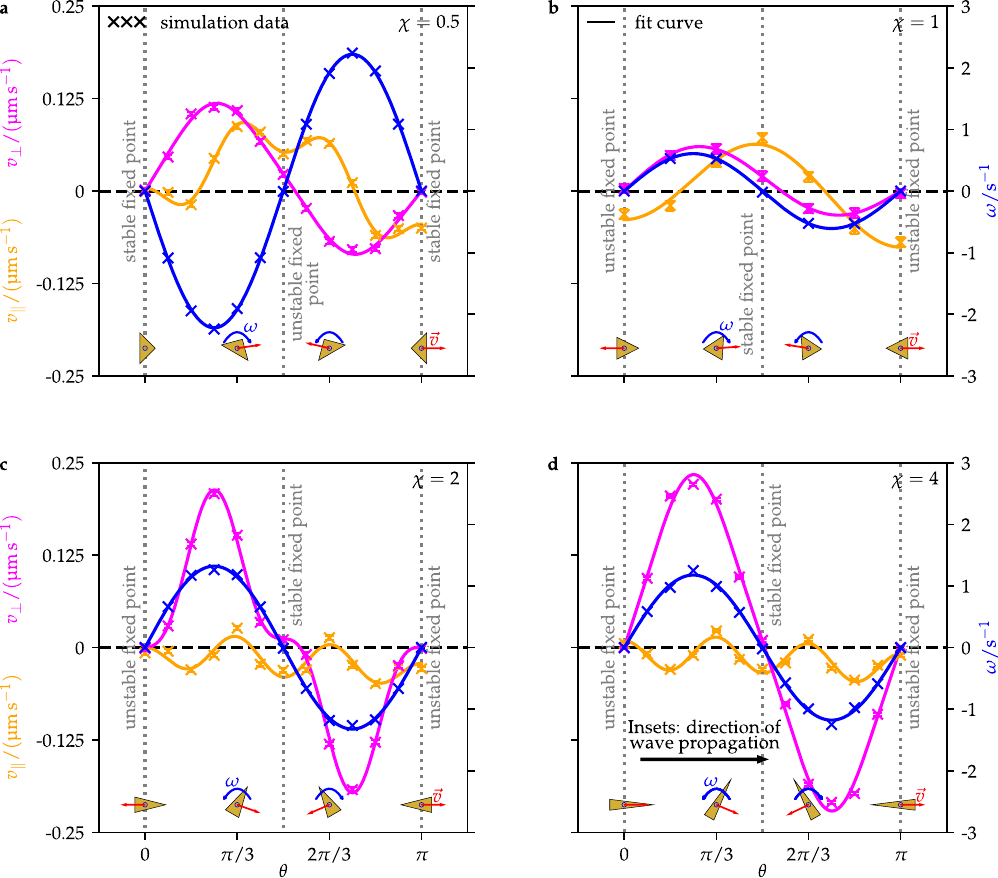}%
\caption{\label{fig:fig3}The particle's translational propulsion velocities $v_\parallel$ (acting along the particle's orientation) and $v_\perp$ (acting perpendicular to the particle's orientation), which are components of the total translational propulsion velocity $\vec{v}=v_\parallel\uu_\parallel + v_\perp\uu_\perp$ of the particle, as well as its angular propulsion velocity $\omega$ are presented as functions of the particle's orientation $\theta\in[ 0,\pi ]$ and for height-to-diameter ratios $\chi=\{0.5, 1, 2, 4\}$ of the particle.}
\end{figure*}

Next, we study how the particle's translational propulsion velocity parallel to the particle's orientation $v_\parallel$, the translational propulsion velocity perpendicular to the particle's orientation $v_\perp$, and the angular propulsion velocity $\omega$ depend on the particle's orientation $\theta$ and aspect ratio $\chi$. 
Our simulation results for $v_\parallel(\theta)$, $v_\perp(\theta)$, and $\omega(\theta)$ for $\chi=0.5,1,2,4$ are shown in Fig.\ \ref{fig:fig3}. Note that only the case $\chi=1$ was previously considered in the literature \cite{VossW2022orientation}. 
\begin{figure*}[tb]
\centering
\includegraphics[width=\linewidth]{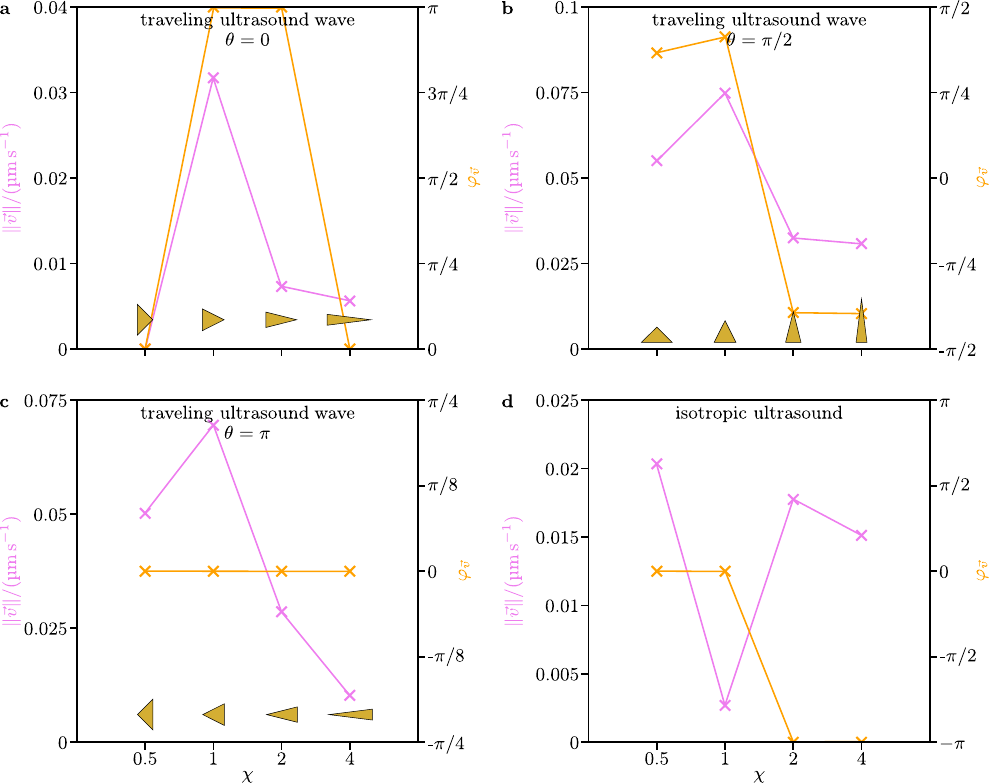}%
\caption{\label{fig4}Dependence of the speed $\norm{\vec{v}}$ and orientation $\varphi_{\vec{v}}$ of the  particles' propulsion on their aspect ratio $\chi$ for a traveling ultrasound wave and (un)stable particle orientations $\theta=0,\pi/2,\pi$ as well as for isotropic ultrasound.}
\end{figure*}
\subsubsection{\label{Description}Description}
For $\chi=0.5$, the parallel velocity $v_\parallel$ is close to zero at $\theta=0$. 
When $\theta$ increases, $v_\parallel$ decreases to a local minimum $v_\parallel=\SI{-0.018}{}\pm\SI{0.003}{\micro\metre\,\second^{-1}}$ at $\theta=\pi/6$, where it starts to increase rapidly to positive values, switching sign between $\theta=\pi/6$ and $\theta=\pi/4$.
After a global maximum $v_\parallel=\SI{0.087}{}\pm\SI{0.003}{\micro\metre\,\second^{-1}}$ at $\theta=\pi/3$, $v_\parallel$ decreases to a local minimum $v_\parallel=\SI{0.050}{}\pm\SI{0.003}{\micro\metre\,\second^{-1}}$ at $\theta=\pi/2$, then it slightly increases again to a local maximum between $\theta=7\pi/12$ and $\theta=2\pi/3$, afterward it strongly decreases, with a sign change between $\theta=3\pi/4$ and $\theta=5\pi/6$, and finally reaches a global minimum $v_\parallel=\SI{-0.060}{}\pm\SI{0.003}{\micro\metre\,\second^{-1}}$ at $\theta=5\pi/6$. 
From there on, $v_\parallel$ increases slowly until $v_\parallel=\SI{-0.050}{}\pm\SI{0.003}{\micro\metre\,\second^{-1}}$ at $\theta=\pi$.

The orientational dependence of the perpendicular velocity $v_\perp$ for $\chi=0.5$ is much simpler. 
It follows a sinus-like function that is zero at $\theta=0$ (this follows from the symmetry properties of the studied system), reaches a global maximum $v_\perp=\SI{0.113}{}\pm\SI{0.003}{\micro\metre\,\second^{-1}}$ at $\theta=\pi/4$, crosses the zero between $\theta=\pi/2$ and $\theta=7\pi/12$, reaches a global minimum $v_\perp=\SI{-0.079}{}\pm\SI{0.003}{\micro\metre\,\second^{-1}}$ at $\theta=3\pi/4$, and becomes zero again at $\theta=\pi$ (for reasons of symmetry).

Also, the function of the angular velocity $\omega$ for $\chi=0.5$ looks relatively simple.
It is similar to a minus-sinus function, where $\omega$ is exactly zero at $\theta=0$ and $\theta=\pi$ (for reasons of symmetry) and close to zero at $\theta=\pi/2$.
The function has a global minimum $\omega=\SI{-2.24}{\second^{-1}}$ at $\theta=\pi/4$ and a global maximum $\omega=\SI{2.24}{\second^{-1}}$ at $\theta=3\pi/4$.
This means that the orientation of the particle has stable fixed points at $\theta=0$ and $\theta=\pi$ and an unstable fixed point near $\theta=\pi/2$. 
The particle will therefore align parallel or antiparallel to the propagation direction of the ultrasound wave.
This finding is interesting since it differs from the behavior of acoustically propelled particles that has been observed in Ref.\ \cite{VossW2022orientation} for a cone-shaped particle with $\chi=1$ and in several experiments \cite{WangCHM2012,AhmedEtAl2013,GarciaGradillaEtAl2013,GarciaGradillaSSKYWGW2014,WangLMAHM2014,AhmedGFM2014,BalkEtAl2014,WuEtAl2015a,Kiristi2015,RaoLMZCW2015,SotoWGGGLKACW2016,AhmedWBGHM2016,EstebanEtAl2016,ZhouYWDW2017,ZhouZWW2017,EstebanFernandezEtAl2017,UygunEtAl2017,XuXZ2017,Zhou2018,SabrinaTABdlCMB2018,WangGWSGXH2018,EstebanEtAl2018,QualliotineEtAl2019,TangEtAl2019,LuSZWPL2019,BhuyanDBSGB2019,DumyJMBGMHA2020,LiuR2020,ValdezLOESSWG2020,McneillSWOLNM2021} for particles with various shapes.

Since $v_\parallel$, $v_\perp$, and $\omega$ are (approximately) zero at $\theta=0$ for $\chi=0.5$ and since this is a stable fixed point for the orientation of the particle, a freely moving particle will reach this state and henceforth be at rest (except for Brownian motion).   
For $\chi=1$, which case has previously been studied in Ref.\ \cite{VossW2022orientation} and is included here only for completeness, the curve for $v_\parallel$ differs from its course that we observed for $\chi=0.5$.
In particular, the curve is now simpler. 
Now, $v_\parallel$ starts with a significantly negative value $v_\parallel=\SI{-0.032}{}\pm\SI{0.007}{\micro\metre\,\second^{-1}}$ at $\theta=0$, increases for increasing $\theta$, with a zero-crossing between $\theta=\pi/6$ and $\theta=\pi/3$, to a global maximum $v_\parallel=\SI{0.072}{}\pm\SI{0.007}{\micro\metre\,\second^{-1}}$ at $\theta=\pi/2$, and from there on decreases, with a zero-crossing between $\theta=2\pi/3$ and $\theta=5\pi/6$, to a global minimum $v_\parallel=\SI{-0.069}{}\pm\SI{0.007}{\micro\metre\,\second^{-1}}$ at $\theta=\pi$. 
The curve for $v_\perp$ is qualitatively similar as for $\chi=0.5$. 
It is zero at $\theta=0$, has a global maximum $v_\perp=\SI{0.057}{}\pm\SI{0.007}{\micro\metre\,\second^{-1}}$ at $\theta=\pi/3$, crosses the zero between $\theta=\pi/2$ and $\theta=2\pi/3$, has a global minimum $v_\perp=\SI{-0.03}{}\pm\SI{0.007}{\micro\metre\,\second^{-1}}$ at $\theta=5\pi/6$, and is zero again at $\theta=\pi$.  
Also, $\omega$ follows a rather simple course again. Compared to the case for $\chi=0.5$, however, the sign of $\omega$ has changed. Now, the curve is zero at $\theta=0$, has a global maximum $\omega=\SI{0.53}{\second^{-1}}$ at $\theta=\pi/6$, crosses the zero close to $\theta=\pi/2$, reaches a global minimum $\omega=\SI{-0.53}{\second^{-1}}$ at $\theta=5\pi/6$, and becomes zero again at $\theta=\pi$.   
Therefore, the particle orientation has now a stable fixed point at $\theta=\pi/2$ and unstable fixed points at $\theta=0$ and $\theta=\pi$, so that the particle will align perpendicular to the direction of propagation of the ultrasound wave, as is well known from experimental studies \cite{WangCHM2012,AhmedEtAl2013,GarciaGradillaEtAl2013,GarciaGradillaSSKYWGW2014,WangLMAHM2014,AhmedGFM2014,BalkEtAl2014,WuEtAl2015a,Kiristi2015,RaoLMZCW2015,SotoWGGGLKACW2016,AhmedWBGHM2016,EstebanEtAl2016,ZhouYWDW2017,ZhouZWW2017,EstebanFernandezEtAl2017,UygunEtAl2017,XuXZ2017,Zhou2018,SabrinaTABdlCMB2018,WangGWSGXH2018,EstebanEtAl2018,QualliotineEtAl2019,TangEtAl2019,LuSZWPL2019,BhuyanDBSGB2019,DumyJMBGMHA2020,LiuR2020,ValdezLOESSWG2020,McneillSWOLNM2021}. 
Hence, the stable and unstable fixed points must switch at some aspect ratio between $\chi=0.5$ and $\chi=1$. 

For $\chi=2$, the curve for $v_\parallel$ starts with $v_\parallel=\SI{-0.007}{}\pm\SI{0.003}{\micro\metre\,\second^{-1}}$ at $\theta=0$ and ends with $v_\parallel=\SI{-0.029}{}\pm\SI{0.003}{\micro\metre\,\second^{-1}}$ at $\theta=\pi$. In between, it oscillates with minima $v_\parallel=\SI{-0.03}{}\pm\SI{0.003}{\micro\metre\,\second^{-1}}$ at $\theta=\pi/6$, $v_\parallel=\SI{-0.031}{}\pm\SI{0.003}{\micro\metre\,\second^{-1}}$ at $\theta=\pi/2$, and $v_\parallel=\SI{-0.049}{}\pm\SI{0.003}{\micro\metre\,\second^{-1}}$ at $\theta=5\pi/6$ and maxima $v_\parallel=\SI{0.026}{}\pm\SI{0.003}{\micro\metre\,\second^{-1}}$ at $\theta=\pi/3$ and $v_\parallel=\SI{0.013}{}\pm\SI{0.003}{\micro\metre\,\second^{-1}}$ at $\theta=2\pi/3$. 
The curve for $v_\perp$ is basically similar to the case $\chi=1$, but the maximum and minimum are now more peaked whereas the curve is now flatter near $\theta=\pi/2$. 
This curve is zero at $\theta=0$, has a global maximum $v_\perp=\SI{0.208}{}\pm\SI{0.003}{\micro\metre\,\second^{-1}}$ at $\theta=\pi/4$, crosses the zero close to $\theta=\pi/2$, reaches a global minimum $v_\perp=\SI{-0.192}{}\pm\SI{0.003}{\micro\metre\,\second^{-1}}$ at $\theta=3\pi/4$, and becomes zero again at $\theta=\pi$.
Moreover, the curve for $\omega$ is qualitatively the same as for $\chi=1$.
Its global maximum is now $\omega=\SI{1.26}{\second^{-1}}$ at $\theta=\pi/4$ and its global minimum is now $\omega=\SI{-1.26}{\second^{-1}}$ at $\theta=3\pi/4$. 
 
For $\chi=4$, the curve for $v_\parallel$ is similar as for $\chi=2$. 
It starts with $v_\parallel=\SI{0.006}{}\pm\SI{0.002}{\micro\metre\,\second^{-1}}$ at $\theta=0$, has minima $v_\parallel=\SI{-0.029}{}\pm\SI{0.002}{\micro\metre\,\second^{-1}}$ at $\theta=\pi/6$, $v_\parallel=\SI{-0.029}{}\pm\SI{0.002}{\micro\metre\,\second^{-1}}$ at $\theta=\pi/2$, and $v_\parallel=\SI{-0.044}{}\pm\SI{0.002}{\micro\metre\,\second^{-1}}$ at $\theta=5\pi/6$, has maxima $v_\parallel=\SI{0.022}{}\pm\SI{0.002}{\micro\metre\,\second^{-1}}$ at $\theta=\pi/3$ and $v_\parallel=\SI{0.011}{}\pm\SI{0.002}{\micro\metre\,\second^{-1}}$ at $\theta=2\pi/3$, and ends with $v_\parallel=\SI{-0.01}{}\pm\SI{0.002}{\micro\metre\,\second^{-1}}$ at $\theta=\pi$. 
Remarkably, the curves for $v_\perp$ and $\omega$ are now quantitatively very similar, where the curve for $\omega$ is quantitatively very similar as for $\chi=2$. 
The curves for $v_\perp$ and $\omega$ follow sinus-like functions that are zero at $\theta=0$ and $\theta=\pi$ and close to zero at $\theta=\pi/2$. Their global maxima are $v_\perp=\SI{0.221}{}\pm\SI{0.002}{\micro\metre\,\second^{-1}}$ and $\omega=\SI{1.25}{\second^{-1}}$ at $\theta=\pi/4$ and their global minima are $v_\perp=\SI{-0.209}{}\pm\SI{0.002}{\micro\metre\,\second^{-1}}$ and $\omega=\SI{-1.25}{\second^{-1}}$ at $\theta=3\pi/4$, respectively.  
 
Based on our results, and taking the findings reported in Refs.\ \cite{VossW2022orientation,WangCHM2012,AhmedEtAl2013,GarciaGradillaEtAl2013,GarciaGradillaSSKYWGW2014,WangLMAHM2014,AhmedGFM2014,BalkEtAl2014,WuEtAl2015a,Kiristi2015,RaoLMZCW2015,SotoWGGGLKACW2016,AhmedWBGHM2016,EstebanEtAl2016,ZhouYWDW2017,ZhouZWW2017,EstebanFernandezEtAl2017,UygunEtAl2017,XuXZ2017,Zhou2018,SabrinaTABdlCMB2018,WangGWSGXH2018,EstebanEtAl2018,QualliotineEtAl2019,TangEtAl2019,LuSZWPL2019,BhuyanDBSGB2019,DumyJMBGMHA2020,LiuR2020,ValdezLOESSWG2020,McneillSWOLNM2021} into account, we can conclude that acoustically propelled particles prefer to orient in such a way that their 
longest axis aligns parallel or antiparallel to the direction of ultrasound propagation. 

Next, we focus on the (un)stable orientations $\theta=0,\pi/2,\pi$ of the particles and study how their speed $\norm{\vec{v}}$ and orientation $\varphi_{\vec{v}}$ (measured analogously to the angle $\theta$) of propulsion depend on $\chi$.
Our corresponding results are shown in Fig.\ \ref{fig4}a-c.
One can see that in each case $\norm{\vec{v}}$ increases from $\chi=0.5$ to a maximum at $\chi=1$ and decreases for larger values of $\chi$.
The behavior of $\varphi_{\vec{v}}$ is qualitatively different for the considered orientations $\theta$. 
For $\theta=0$, $\varphi_{\vec{v}}$ is $0$ or $\pi$, i.e., the propulsion is parallel or antiparallel to the particle's orientation, depending on the value of $\chi$. 
In the case $\theta=\pi/2$, $\varphi_{\vec{v}}$ is between $-\pi/2$ and $-\pi/4$ or between $\pi/4$ and $\pi/2$ depending on the value of $\chi$. 
Interestingly, for $\theta=\pi$, we observe $\varphi_{\vec{v}}=0$ for all values of $\chi$, i.e., the propulsion is always antiparallel to the orientation of the particle.

\subsubsection{\label{AnalyticRepresentation}Analytic representation}
To help readers of this article to build upon our work, we present also an analytic representation of our simulation results for $v_\parallel$, $v_\perp$, and $\omega$.
The analytic representation consists of fit functions for the velocities $v_\parallel(\theta)$, $v_\perp(\theta)$, and $\omega(\theta)$ for each considered value of the aspect ratio $\chi$. 

We first consider the properties of the functions $v_\parallel(\theta)$, $v_\perp(\theta)$, and $\omega(\theta)$.
From the setup of the studied system, we can infer the symmetry properties 
\begin{align}
v_\parallel(-\theta) &= v_\parallel(\theta), \\
v_\perp(-\theta) &= -v_\perp(\theta), \\
\omega(-\theta) &= -\omega(\theta)
\end{align}
and the side conditions
\begin{align}
v_\perp(0) &= v_\perp(\pi)=0, \\
\omega(0) &= \omega(\pi)=0. 
\end{align}
Taking these features of the functions $v_\parallel(\theta)$, $v_\perp(\theta)$, and $\omega(\theta)$ into account, a Fourier series ansatz 
\begin{align}
v_\parallel(\theta) &= a_{\parallel,0} + a_{\parallel,1}\cos(\theta) 
+ \sum_{i=1}^{3} a_{\parallel,i+1} \cos(2i\theta), \label{eq:fit_parallel} \\
\begin{split}
v_\perp(\theta) &= a_{\perp,1}\sin(\theta) \\ 
&\quad\,+ \sum_{i=1}^{2} a_{\perp,i+1}\sin((2i-1)\theta)\cos((2i-1)\theta), \label{eq:fit_perp} 
\end{split}\\
\omega(\theta) &= a_{\omega} \sin(2\theta) \label{eq:fit_omega}
\end{align}
can be used as an analytic representation of the functions. 
As a consequence of the rather simple curves that we observed for $v_\parallel(\theta)$, $v_\perp(\theta)$, and $\omega(\theta)$ in Section \ref{Description}, a Fourier series ansatz of this low order is already sufficient to reach a very good agreement of the simulation data and the analytic fit curves. 

By fitting the functions \eqref{eq:fit_parallel}-\eqref{eq:fit_omega} to our simulation data, we obtained the values of the fit coefficients $a_{\parallel,0},\dotsc,a_{\parallel,4}$, $a_{\perp,1},a_{\perp,2},a_{\perp,3}$, and $a_{\omega}$ that are listed in Tab.\ \ref{tab:Fitparameter}.  
\begin{table*}[tb]
\caption{\label{tab:Fitparameter}Values of the fit coefficients of the functions \eqref{eq:fit_parallel}-\eqref{eq:fit_omega}, corresponding to the velocity components $v_\parallel$, $v_\perp$, and $\omega$, for the particle's aspect ratios $\chi=0.5,1,2,4$.}
\centering
\begin{ruledtabular}
\begin{tabular}{ccccccc}
\textbf{Velocity} & $\boldsymbol{\chi}$ & $\boldsymbol{a_{\parallel,0}}$ or $\boldsymbol{a_{\omega}}$ & $\boldsymbol{a_{\parallel,1}}$ or $\boldsymbol{a_{\perp,1}}$ & $\boldsymbol{a_{\parallel,2}}$ or $\boldsymbol{a_{\perp,2}}$ & $\boldsymbol{a_{\parallel,3}}$ or $\boldsymbol{a_{\perp,3}}$ & $\boldsymbol{a_{\parallel,4}}$\\
\hline
$v_\parallel$& $\SI{0.5}{} $ & $ \SI{2.035}{\cdot10^{-2}\,\micro\metre\,\second^{-1}}$ & $\SI{2.434}{\cdot10^{-2}\,\micro\metre\,\second^{-1}} $ & $\SI{-6.158}{\cdot10^{-2}\,\micro\metre\,\second^{-1}} $ & $\SI{-4.119}{\cdot10^{-3}\,\micro\metre\,\second^{-1}}$ & $\SI{2.487}{\cdot10^{-2}\,\micro\metre\,\second^{-1}}$ \\
$v_\perp$  & $\SI{0.5}{}$ &  ---  & $ \SI{2.346}{\cdot10^{-2}\,\micro\metre\,\second^{-1}}$ & $\SI{1.961}{\cdot10^{-1}\,\micro\metre\,\second^{-1}} $ & $ \SI{-7.505}{\cdot10^{-3}\,\micro\metre\,\second^{-1}} $ &  --- \\
$\omega$ & $\SI{0.5}{}$ & $\SI{-2.22}{\second^{-1}}$ & --- & --- & --- &  --- \\
$v_\parallel$& $\SI{1}{}$ & $ \SI{2.691}{\cdot10^{-3}\,\micro\metre\,\second^{-1}}$ & $\SI{1.864}{\cdot10^{-2}\,\micro\metre\,\second^{-1}} $ & $\SI{-5.999}{\cdot10^{-2}\,\micro\metre\,\second^{-1}} $ & --- & --- \\
$v_\perp$  & $\SI{1}{}$ &  ---  & $ \SI{1.949}{\cdot10^{-2}\,\micro\metre\,\second^{-1}}$ & $\SI{9.169}{\cdot10^{-2}\,\micro\metre\,\second^{-1}} $ & --- &  --- \\
$\omega$ & $\SI{1}{}$ & $\SI{6.079}{\cdot10^{-1}\,\second^{-1}}$ & --- & --- & --- &  --- \\
$v_\parallel$& $\SI{2}{}$ & $ \SI{-1.775}{\cdot10^{-2}\,\micro\metre\,\second^{-1}}$ & $\SI{1.085}{\cdot10^{-2}\,\micro\metre\,\second^{-1}} $ & $\SI{-5.826}{\cdot10^{-3}\,\micro\metre\,\second^{-1}} $ & $\SI{-7.376}{\cdot10^{-3}\,\micro\metre\,\second^{-1}}$ & $\SI{2.06}{\cdot10^{-2}\,\micro\metre\,\second^{-1}}$ \\
$v_\perp$  & $\SI{2}{}$ &  ---  & $ \SI{1.198}{\cdot10^{-2}\,\micro\metre\,\second^{-1}}$ & $\SI{3.085}{\cdot10^{-1}\,\micro\metre\,\second^{-1}} $ & $ \SI{-1.009}{\cdot10^{-1}\,\micro\metre\,\second^{-1}} $ &  --- \\
$\omega$ & $\SI{2}{}$ & $\SI{1.322}{\second^{-1}}$ & --- & --- & --- &  --- \\
$v_\parallel$& $\SI{4}{}$ & $ \SI{-1.512}{\cdot10^{-2}\,\micro\metre\,\second^{-1}}$ & $\SI{8.91}{\cdot10^{-3}\,\micro\metre\,\second^{-1}} $ & $\SI{-4.319}{\cdot10^{-3}\,\micro\metre\,\second^{-1}} $ & $\SI{-1.81}{\cdot10^{-3}\,\micro\metre\,\second^{-1}}$ & $\SI{2.183}{\cdot10^{-2}\,\micro\metre\,\second^{-1}}$ \\
$v_\perp$ & $\SI{4}{}$ & ---  & $ \SI{9.188}{\cdot10^{-3}\,\micro\metre\,\second^{-1}}$ & $\SI{4.3}{\cdot10^{-1}\,\micro\metre\,\second^{-1}} $ & $ \SI{-2.481}{\cdot10^{-2}\,\micro\metre\,\second^{-1}} $ &  --- \\
$\omega$ & $\SI{4}{}$ & $\SI{1.178}{\second^{-1}}$ & --- & --- & --- &  --- 
\end{tabular} 
\end{ruledtabular}%
\end{table*}
As is evident from Fig.\ \ref{fig:fig3}, the agreement of the fit functions with our simulation data is very good for these coefficient values.

\subsection{Dependence of orientation-averaged propulsion on aspect ratio}
In future applications of acoustically propelled particles, the particles might also be deployed in isotropic ultrasound fields instead of a traveling ultrasound wave. 
To assess, how the particles that are studied in the present work would behave when they are exposed to isotropic ultrasound, we calculate their corresponding propulsion velocities by averaging the functions $v_\parallel(\theta)$, $v_\perp(\theta)$, and $\omega(\theta)$ over the orientation $\theta\in [-\pi,\pi)$.
Using the analytic representation of these functions from Section \ref{AnalyticRepresentation}, we obtain the orientation-averaged propulsion velocities
\begin{align}
\langle v_\parallel(\theta)\rangle_\theta &= a_{\parallel,0}, \\
\langle v_\perp(\theta)\rangle_\theta &= 0,\\
\langle\omega(\theta)\rangle_\theta &= 0
\end{align}
with the angular average
\begin{equation}
\langle W(\theta) \rangle_\theta = \frac{1}{2\pi} \int_{-\pi}^{\pi} \!\!\!\! W(\theta)  \,\mathrm{d}\theta ,
\end{equation}
where $W(\theta)$ is a wildcard function. 

We thus see that in isotropic ultrasound all particles show purely translational propulsion parallel or antiparallel to the instantaneous particle orientation. 
Furthermore, using the fit values listed in Tab.\ \ref{tab:Fitparameter}, we can identify the particle with aspect ratio $\chi=0.5$ as the particle with the fastest propulsion in isotropic ultrasound.
For the acoustic energy density $E=\SI{22.7}{\milli\joule\,\metre^{-3}}$ that is used in our simulations, we obtain the orientationally averaged parallel propulsion velocity $\langle v_\parallel\rangle_\theta\approx\SI{0.02}{\micro\metre\,\second^{-1}}$. 
This is roughly 10 times larger than the orientationally averaged parallel propulsion velocity that has previously been reported for $\chi=1$ \cite{VossW2022orientation}. 
Furthermore, the value $\langle v_\parallel\rangle_\theta\approx\SI{0.02}{\micro\metre\,\second^{-1}}$ is significantly greater than the approximate numerical error $\SI{0.003}{\micro\metre\,\second^{-1}}$. Note that this was not the case for the value of $\langle v_\parallel\rangle_\theta$ that was previously reported for $\chi=1$ in Ref.\ \cite{VossW2022orientation}. 
The dependence of the propulsion in isotropic ultrasound on the aspect ratio $\chi$ is visualized in Fig.\ \ref{fig4}d.
This figure shows that the propulsion speed is maximal for $\chi=0.5$, minimal for $\chi=1$, and moderate for larger $\chi$. While the propulsion is parallel to the particle's orientation for $\chi=0.5$ and $\chi=1$, the propulsion direction and particle orientation are antiparallel for the larger values of $\chi$.
Thus, the dependence of the propulsion on $\chi$ is qualitatively different for particles in isotropic ultrasound than for particles in a traveling ultrasound wave, where the propulsion speed is maximal for $\chi=1$ (see Section \ref{Description}). 

Since the propulsion velocity is approximately proportional to the acoustic energy density $E$ \cite{VossW2022acoustica,VossW2021}, we can easily determine the particle's orientationally averaged parallel propulsion velocity for larger values of $E$. 
With respect to future applications of acoustically propelled particles in nanomedicine, the energy density $E_\mathrm{max}=\SI{4.9}{\joule \metre^{-3}}$, which constitutes an upper limit for diagnostic applications of ultrasound \cite{BarnettEtAl2000}, is particularly relevant. 
Rescaling the particle's orientationally averaged parallel propulsion velocity for $\chi=0.5$ according to this larger acoustic energy density, we obtain the orientationally averaged parallel propulsion velocity 
$\langle v_\parallel\rangle_{\theta,\mathrm{rescaled}} \approx \SI{4.4}{\micro\metre\,\second^{-1}}$, which equals roughly a speed of 4 times the particle size per second. 
Such a speed should be sufficient for a number of medical or technical future applications of such particles. 
Note that, as we have seen further above, the particles reach, for a given acoustic energy density $E$, much larger propulsion speeds in a traveling ultrasound wave than in isotropic ultrasound.

\section{\label{conclusions}Conclusions}
We have studied the orientation-dependent acoustic propulsion and associated time-averaged flow fields of cone-shaped particles with different aspect ratios by a traveling planar ultrasound wave as well as the particles' propulsion by isotropic ultrasound.
Knowing the propulsion of such particles is crucial with respect to the envisaged future application of ultrasound-propelled particles in fields like nanomedicine \cite{LiEFdAGZW2017,PengTW2017,SotoC2018,WangGZLH2020,WangZ2021} and materials science \cite{JunH2010,McdermottKDKWSV2012,KuemmelSLVB2015,VanderMeerDF2016,JeanneretPKP2016,NeedlemanD2017,WangDZLGYLWM2019,RamananarivoDP2019}.
While current studies predominantly investigate the motion of nano- and microparticles in a standing ultrasound wave \cite{WangCHM2012,AhmedEtAl2013,GarciaGradillaEtAl2013,GarciaGradillaSSKYWGW2014,WangLMAHM2014,AhmedGFM2014,BalkEtAl2014,WuEtAl2015a,EstebanFernandezdeAvilaMSLRCVMGZW2015,XuEtAl2015,Kiristi2015,RaoLMZCW2015,SotoWGGGLKACW2016,AhmedWBGHM2016,EstebanEtAl2016,ZhouYWDW2017,ZhouZWW2017,EstebanFernandezEtAl2017,UygunEtAl2017,XuXZ2017,Zhou2018,SabrinaTABdlCMB2018,WangGWSGXH2018,EstebanEtAl2018,QualliotineEtAl2019,TangEtAl2019,LuSZWPL2019,BhuyanDBSGB2019,DumyJMBGMHA2020,LiuR2020,ValdezLOESSWG2020,McneillSWOLNM2021}, in future applications the particles will more likely be exposed to a traveling ultrasound wave or to isotropic ultrasound \cite{VossW2022orientation}.
The fact that we have observed a strong dependence of the acoustic propulsion on the particle orientation relative to the propagation direction of the ultrasound wave for all aspect ratios of the particle shows that it is not sufficient to study acoustically propelled particles in a standing ultrasound wave. 
Therefore, future research should focus more on application-relevant setups and continue this work, e.g., by studying the orientation-dependent propulsion of further particles with other shapes. 

A remarkable finding of this work is that, for all considered aspect ratios, a cone-shaped particle tends to align with a certain angle relative to the propagation direction of the ultrasound, and that this alignment changes from parallel or antiparallel alignment for small aspect ratios to perpendicular alignment for larger aspect ratios. 
A similar effect could occur for other particle designs that are not studied in the present work.  
This shows that in the typical experimental setups, where the motion of particles is observed in a nodal plane of a standing ultrasound wave \cite{WangCHM2012,AhmedEtAl2013,GarciaGradillaEtAl2013,GarciaGradillaSSKYWGW2014,WangLMAHM2014,AhmedGFM2014,BalkEtAl2014,WuEtAl2015a,Kiristi2015,RaoLMZCW2015,SotoWGGGLKACW2016,AhmedWBGHM2016,EstebanEtAl2016,ZhouYWDW2017,ZhouZWW2017,EstebanFernandezEtAl2017,UygunEtAl2017,XuXZ2017,Zhou2018,SabrinaTABdlCMB2018,WangGWSGXH2018,EstebanEtAl2018,QualliotineEtAl2019,TangEtAl2019,LuSZWPL2019,BhuyanDBSGB2019,DumyJMBGMHA2020,LiuR2020,ValdezLOESSWG2020,McneillSWOLNM2021} in which the particles levitate, some particle designs, such as cone-shaped particles with a small aspect ratio, can in principle have efficient acoustic propulsion but might not show significant propulsion in the nodal plane since they align parallel or antiparallel to the standing ultrasound wave and thus perpendicular to the nodal plane.   
On the other hand, this effect will provide an interesting ansatz for guiding the motion of acoustically propelled particles.  
Furthermore, the strong dependence of the alignment angle on the aspect ratio of the particles allows to utilize this effect for sorting particles by their shape. 

The observation that cone-shaped particles with aspect ratio $\chi=0.5$ have a stable state where they show no translational or angular acoustic propulsion suggests to reserve the application of particles with this particular aspect ratio for special purposes. For typical envisaged applications such as drug delivery, these particles are rather inappropriate, since whenever a particle reaches an orientation parallel to the direction of ultrasound propagation the particle will stop moving and remain in this state until it is reoriented by a sufficient angle through Brownian rotation or external torques. 

Finally, it is likely that the provided analytic representation of our simulation results will prove helpful for future studies. For example, these analytic expressions allow to incorporate the acoustic propulsion of a particle in a relatively simple way into a particle-based \cite{WittkowskiL2012,tenHagenWTKBL2015} or field-based \cite{BickmannW2020twoD,BickmannW2020b,teVrugtLW2020} model for acoustically propelled particles, where the characteristic time scale of the model can be many orders of magnitude larger than the time period of the ultrasound wave and an explicit description of the ultrasound propagation is not necessary.

\section*{Data availability}
The raw data corresponding to the figures shown in this article are available as Supplementary Material at \url{https://doi.org/10.5281/zenodo.5947177}.

\section*{Conflicts of interest}
There are no conflicts of interest to declare.

\begin{acknowledgments}
We thank Patrick Kurzeja for helpful discussions. 
R.W.\ is funded by the Deutsche Forschungsgemeinschaft (DFG, German Research Foundation) -- 283183152 (WI 4170/3).
The simulations for this work were performed on the computer cluster PALMA II of the University of M\"unster. 
\end{acknowledgments}

\nocite{apsrev41Control}
\bibliographystyle{apsrev4-1}
\bibliography{control,refs}
	
\end{document}